\documentclass{svjour3}

\usepackage[utf8]{inputenc}
\usepackage{graphicx}
\usepackage{url}
\usepackage{natbib}

\smartqed

\journalname{arXiv}

\title{Quality Assessment and Improvement of Helm Charts for Kubernetes-Based Cloud Applications}
\titlerunning{Quality Assessment and Improvement of Helm Charts for ... Cloud Applications}
\author{Josef Spillner}

\institute{J. Spillner \at
Zurich University of Applied Sciences, School of Engineering\\
Service Prototyping Lab (blog.zhaw.ch/splab), Winterthur, Switzerland\\
Tel.: +41 58 934 45 82\\
\email{josef.spillner@zhaw.ch}\\
ORCID: 0000-0002-5312-5996}

\date{\today}

\begin{document}

\maketitle

\begin{abstract}
Helm has recently been proposed by practitioners as technology to package
and deploy complex software applications on top of Kubernetes-based cloud computing platforms.
Despite growing popularity, little is known about the individual so-called \textit{Helm Charts}
and about the emerging ecosystem of charts around the \textit{KubeApps Hub} website and
decentralised charts repositories.
This article contributes first quantified insights around both the charts and the artefact development community
based on metrics automatically gathered by a proposed quality assessment tool named \textit{HelmQA}.
The work further identifies quality insufficiencies detectable in public charts, proposes
a developer-centric hypothesis-based methodology to systematically improve the quality by using HelmQA,
and finally empirically attempts to validate the methodology and thus the practical usefulness of the tool
by presenting results of its application over a representative four-month period.
Although one of our initial hypotheses does not statistically hold during the experiment, we still infer
that using HelmQA regularly in continuous software development would lead to reduced quality issues.

\end{abstract}

\begin{acknowledgements}
This research has been partially funded by Innosuisse - Swiss Innovation Agency in project MOSAIC/19333.1.
\end{acknowledgements}

\section{Introduction}

Digital ecosystems built around artefact repositories are important cornerstones for global-scale service delivery in which
software developers, platform operators and service end users are often assuming overlapping roles.
Cloud computing has emerged as one of the principal computing paradigms to build such ecosystems
and therefore bring advances to society through collaborative work \cite{DBLP:journals/cloudcomp/Jeffery18}.

On a technological level, innovations in cloud computing are driven by two main forces.
Singular commercial cloud service providers which regularly create unique selling points for their offerings \cite{DBLP:journals/ibmrd/ArnoldBDKMNSSSS16},
and open source stacks which through multiplicity in various deployments and commercial enhancements achieve
a broad impact and a high de-facto standardisation and research enablement potential \cite{DBLP:conf/ccgrid/CherrueauPSLS17}.
In recent years, such stacks have included OpenStack, OpenNebula, Docker and Kubernetes with correspondingly high
numbers of related publications, often with direct practical impact \cite{DBLP:conf/edcc/EliaALV17,DBLP:conf/gecon/GraciaTATBR17,DBLP:conf/fast/AnwarMTLRCZNWLH18}.
In these frameworks, services are increasingly
composed of smaller executable software entities which are choreographed and orchestrated into larger
units.

The trend towards the construction of composite applications based on microservices has led to the need
to represent the composition itself as tangible entity. In analogy to popular mobile phone apps,
the compositions would appear as user-facing cloud apps.
Previous approaches have included executable business processes,
container composition files and sets of deployment descriptor files.
All three approaches have proven to be limited concerning the tangible representation.

Business process execution in the XML format of BPEL (Business Process Execution Language) merely defines interaction
protocols of available services without defining how to make the services available \cite{DBLP:conf/iisa/RouisBSK17},
a severe gap given the requirements of service placement, service replication and adaptive decisions about dependency services.
Container composition files in the Docker Compose format cover the instantiation of containers but do not directly support
service semantics. Containers may offer zero, one or more services. Furthermore, resource allocation constraints are not
considered in this format.
Descriptor files in the Kubernetes format mitigate this limitation but due to a multitude of deployment
and service descriptors lead to redundant values and more difficult handling. Thus, Kubernetes apps would require
a higher-level treatment.

To overcome the limitation in Kubernetes stacks, Helm has been proposed in mid-2016 by practitioners as solution to
package sets of descriptor files, including templates and rich metadata, into single archive files which can be deployed and undeployed
easily. Helm therefore defines a packaging file format, called \textit{Helm Charts}, as well as client and server components to handle
such files. On the server side, the Tiller implementation is deployed on top of Kubernetes, and on the client side,
the Helm binary allows for creation, testing and deployment of charts as well as searching across repositories.
The format of Helm charts is specified informally through an evolving technology
documentation\footnote{Helm charts format: \url{https://docs.helm.sh/developing_charts/}}.

Considering that Helm charts increasingly form the building blocks of complex cloud software, insisting strictly on
their quality is essential to avoid a propagation of the quality issues into the software development process.
Helm already defines a \texttt{lint} command to check for quality issues in charts. However, the linting is limited
to well-formedness. According to quality standards, well-formedness is a necessary although insufficient prerequisite
for validity and furthermore does not imply that format expressions are optimal. Moreover, the linting is confined to
single charts and does not perform cross-package checks which are essential as soon as dependencies are
involved \cite{DBLP:journals/ese/CaneillGZ17}.
Hence, a deeper analytical inspection of Helm charts and potential issues is required.

In this paper, we contribute the first such analysis along with enabling quality assessment software.
Specifically, we formalise aspects of Helm charts, provide
insights into the current Helm ecosystem as well as the quality of individual charts, quantify and visualise associated artefact and community metrics,
and identify insufficiencies. Moreover, we report on our empirical approach to improve the charts quality over a longer period
of time. We claim that the integration of our quality checks into the \texttt{lint} command and into the chart submission process
would lead to overall improvements of present and future Helm charts. To enable this evolution, we provide reference checks as open source software
called HelmQA which integrates into the development workflow for cloud apps.

\section{Analysis}

We first study the \textit{KubeApps Hub} repository which is the official site to retrieve Helm charts,
both as a user-facing website and as a programmable endpoint used by the Helm implementation.
Helm itself can be considered a fairly mature implementation with two years of development time and
around 4400 code commits, which is however relative to Kubernetes' 70000 commits over an only slightly longer
period of time. On the other hand, the hub has emerged later and has received around 670 commits since October 2017
which implies an active albeit ongoing implementation\footnote{Based on GitHub projects: kubernetes/helm, kubernetes/kubernetes, kubeapps/kubeapps}.

\begin{figure}[h]
\centering
\caption{Entity-relationship diagram for KubeApps Hub}
\label{fig:kubeappshub}
\includegraphics[width=0.7\columnwidth]{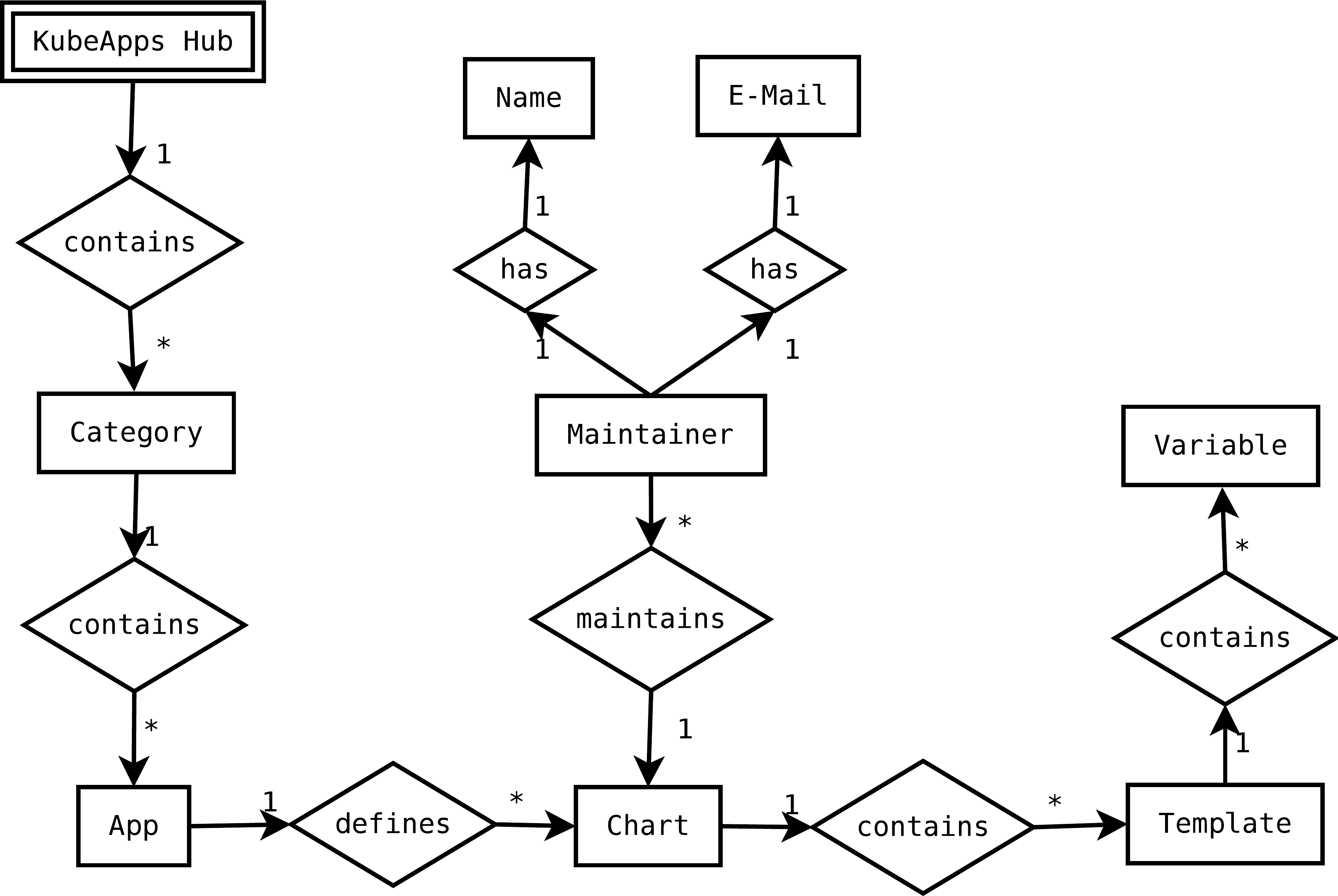}
\end{figure}

The hub repository is divided into a \texttt{stable} and an \texttt{incubator} section which are subdivided
into \textit{apps} with technical representation as charts.
Fig. \ref{fig:kubeappshub} formally defines the entities and relationships for digital objects of interest
on KubeApps Hub. This entity-relationship model serves as basis for all following analysis steps.

A snapshot of May 15 2018, the starting point of our study, reveals the presence of 188 apps overall. Among them are
153 stable apps which are of interest to our study as well as 35 incubator apps which, due to frequent
changes and trials, will be discarded from our preliminary analysis. Due to occasional multi-versioning, 177 stable charts are present,
resulting in a versioning overhead of 15.7\%.

Evidently, not all charts are registered at KubeApps Hub. Owing to the popularity of GitHub, we include it as second
data point to complement the view on charts with decentralised and less quality-controlled charts. A custom search for
charts in May 2018 yields 24560 hits. Due to search API restrictions, only the first 1000 of those are sampled. They are distributed
across 572 repositories, giving an average of 1.75 charts and a maximum of 31 charts per repository.
A further sampling takes place for a detailed analysis in which all repositories containing at least 10 charts as threshold
value are included, giving a subset of 7 repositories and 134 hits. Due to test files and other sources of imprecision, eventually
a set of 115 distinct charts is contained in the subset. Any overlap with KubeApps Hub has not been checked but
can be considered insignificant; rather, many charts found in GitHub can be considered forks of others which implies
a much smaller number of stem charts.

We make use of visual directed graph representations as well as scripted quality checks to achieve quantified analytical
results which will be outlined in the next paragraphs based on the mid-May 2018 data acquisition.
Many of the graphs included in this paper are directly produced by the scripts. Along with reference results they
are available as re-usable open source software package named
\textit{HelmQA}\footnote{HelmQA code repository: \url{https://github.com/serviceprototypinglab/helmqa}}.

\subsection{Maintenance insights}

Helm charts consist of a meta information file (\texttt{Chart.yaml}), a directory of Kubernetes deployment
descriptors supporting a template syntax for value substitution (\texttt{templates/}), a single file with substitution
values (\texttt{values.yaml}) as well as dependency information (\texttt{requirements.txt}, \texttt{charts/})
Templates are descriptive YAML files mixed with imperative Go conditional branching and formatting directives which can be rendered into
valid YAML descriptors by substituting the placeholders for values from \texttt{values.yaml}.

\begin{figure}[h]
\centering
\caption{Diverse (n:1, 1:n) maintainer-chart relationships}
\label{fig:excerpt2}
\includegraphics[width=0.873\columnwidth]{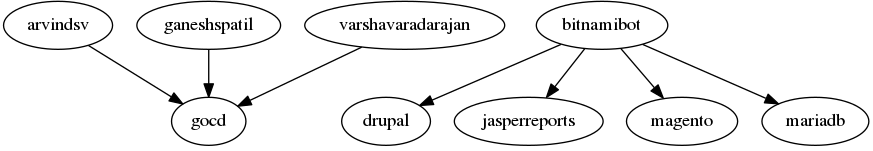}
\end{figure}

The meta file lists among other information the maintainers of a chart which are typically developers or adressable
teams. Charts can be co-maintained by multiple developers. Maintainers are listed with a combination of name, by convention
in the sense of short username, and e-mail address.
Fig. \ref{fig:excerpt2} shows typical constallations resulting from this rule.
As representatives of the n:m relationship between charts and maintainers, the chart \texttt{gocd} has three maintainers
whereas the team account \texttt{bitnamibot} maintains 14 charts of which four are shown.

\begin{figure}[h]
\centering
\caption{Complex (transitive) maintainer-chart relationships}
\label{fig:excerpt}
\includegraphics[width=0.784\columnwidth]{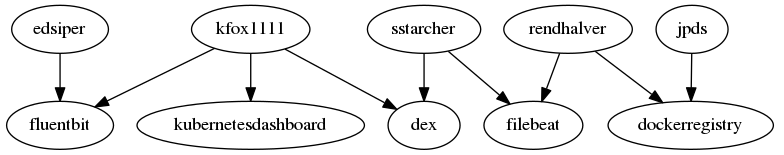}
\end{figure}

More complex relationships are possible as evidenced by Fig. \ref{fig:excerpt}. Developers \texttt{kfox1111} and \texttt{sstarcher}
jointly co-maintain one chart, \texttt{dex}, while individually or with other co-maintainers caring about other
charts. This constellation suggests a high degree of collaborative and community-driven work.

\begin{figure}[h]
\centering
\caption{Irregularity due to multiple stable versions of the same chart (left) and due to the equality between chart and maintainer name (right)}
\label{fig:excerpt34}
\includegraphics[width=0.308\columnwidth]{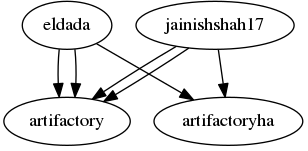}
\hspace{4.0mm}
\includegraphics[width=0.303\columnwidth]{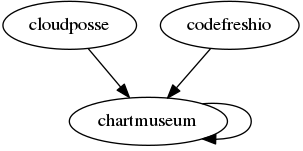}
\end{figure}

Irregularities occur in various forms. For instance, double arrows occur when more than one version
of a chart is found within the same index, in this case the stable category of KubeApps Hub,
and self-references occur for collisions when a chart name matches the maintainer name (Fig. \ref{fig:excerpt34} for \texttt{artifactory}
and \texttt{chartmuseum} respectively). While not prohibited by the Helm specification,
such irregularities may cause confusion among developers and may combine with more harmful issues.

\begin{figure}[h]
\centering
\caption{Irregularity due to maintainer alias names (left) and due to absence of maintainers (right)}
\label{fig:excerpt56}
\includegraphics[width=0.440\columnwidth]{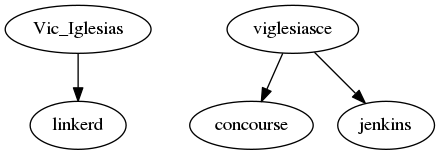}
\hspace{4.0mm}
\includegraphics[width=0.286\columnwidth]{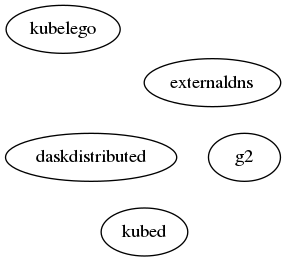}
\end{figure}

More irregularities occur when no maintainer is defined or when, across packages, maintainers
have multiple identities, referring to multiple names linked to the same e-mail address (Fig. \ref{fig:excerpt56}).
The latter irregularity includes the absence of an e-mail address.

Apart from the irregularities, differences occur in the use of the templating mechanism. Some charts
make excessive use of templates to reduce redundant values in the deployment descriptors while others
use them lightly. A semantic assessment of the right level of templating is not trivial, but the level
of remaining redundancy can be quantified, fed into a decision-making process and subsequently consciously
minimised.

Potential benefits of a systematic quality assurance of charts before their provisioning on repositories
are evident, yet adequate tools are sparse. The Helm \texttt{lint} command, the primary quality assurance instrument
for chart developers, is described as follows:
\textit{This command takes a path to a chart and runs a series of tests to verify that the chart is well-formed.}
According to the KubeApps Hub rules, a chart must pass this command in order to be included in the stable
category, and yet as shown several irregularities remain without being pointed out by the tool.

\subsection{Chart and maintenance metrics}

Several metrics quantify the initial state of all stable Helm charts at KubeApps Hub as well as
the control set on GitHub.
All of the following metrics and plots have been produced by the HelmQA scripts and can thus
be reproduced.

\begin{figure}[h]
\centering
\caption{Maintainers and charts per maintainer set on KubeApps Hub}
\label{fig:cross}
\includegraphics[width=0.8\columnwidth]{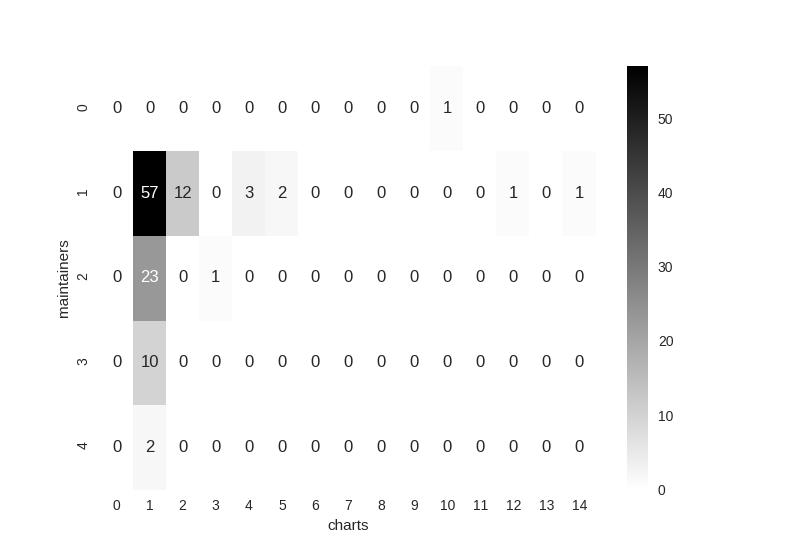}
\end{figure}

Fig. \ref{fig:cross} presents a heatmap view on the KubeApps Hub relation between maintainers and maintained
charts per maintainer set for a total of 113 sets.
In 57 cases, a set consists of a single maintainer for a single chart (50.44\%).
Other noteworthy cases are single chart maintenance by two or three maintainers (combined 33 or 29.20\%),
two charts maintained by a single person (12 or 10.62\%), and charts apparently maintained
by nobody (10 in total represented as single set or 0.89\%).
Overall, not considering the charts without maintainer information,
129 charts (72.88\%) are maintained alone and 38 charts (21.47\%) are maintained collaboratively.
For comparison, 82.61\% of charts in the GitHub sample do not contain maintainer metadata which
demonstrates a lack of quality assurance in the fully decentralised development model.

Table \ref{tab:maintainer} lists maintainer-specific metrics such as the composition and
workload of teams of maintainers further described as maintainer sets on KubeApps Hub. It includes
one irregularity observed in maintainer information.
The corresponding metrics from the GitHub sample are listed in Table \ref{tab:maintainergithub}.
The higher number of charts is explained by the choice of sampling and the high number of charts without maintainer information
whereas the maintainer set sizes are comparable. However, the irregularity in alias names is less severe which indicates
that despite some required quality checks, KubeApps Hub metrics are not consistently better.

\begin{table}[h]
\centering
\caption{Maintainer-related KubeApps Hub metrics}
\label{tab:maintainer}
\begin{tabular}{lrrl}\hline
Maintainer metrics		& Affected	& Percentage	& Base metric	\\ \hline

Maintainers			& 142		& 100.00	& --		\\
Sets of maintainers		& 113		& 100.00	& --		\\
Avg charts per maintainer	& 1.25		& 0.71		& Charts	\\
Avg charts per maintainer set	& 1.57		& 8.87		& -""-		\\
Max charts per maintainer set	& 14		& 7.91		& -""-		\\
Avg maintainers per set		& 1.26		& 0.89		& Maintainers	\\
Max maintainers per set		& 4		& 2.82		& -""-		\\
Irregularity: alias names	& 8		& 6.72		& Unique e-mails	\\ \hline
\end{tabular}
\end{table}

\begin{table}[h]
\centering
\caption{Maintainer-related GitHub metrics}
\label{tab:maintainergithub}
\begin{tabular}{lrrl}\hline
Maintainer metrics		& Affected	& Percentage	& Base metric	\\ \hline

Maintainers			& 19		& 100.00	& --		\\
Sets of maintainers		& 14		& 100.00	& --		\\
Avg charts per maintainer	& 6.05		& 5.26		& Charts	\\
Avg charts per maintainer set	& 8.21		& 7.14		& -""-		\\
Max charts per maintainer set	& 4		& 3.48		& -""-		\\
Avg maintainers per set		& 1.35		& 7.11		& Maintainers	\\
Max maintainers per set		& 3		& 15.8		& -""-		\\
Irregularity: alias names	& 1		& 5.56		& Unique e-mails	\\ \hline
\end{tabular}
\end{table}

Tables \ref{tab:chart} and \ref{tab:chartgithub} complement these with chart-specific views including
quantified irregularities in the chart descriptions. Again, only some metrics indicate higher quality
in KubeApps Hub.

\begin{table}[h]
\centering
\caption{Chart-related KubeApps Hub metrics}
\label{tab:chart}
\begin{tabular}{lrrl}\hline
Chart metrics			& Affected	& Percentage	& Base metric	\\ \hline

Charts				& 177		& 100.00	& --		\\
Irregularity: no maintainer	& 10		& 5.65		& Charts	\\
Irregularity: name collision	& 1		& 0.56		& Charts	\\
Irregularity: multiple versions	& 5		& 2.82		& Charts	\\ \hline
\end{tabular}
\end{table}

\begin{table}[h]
\centering
\caption{Chart-related GitHub metrics}
\label{tab:chartgithub}
\begin{tabular}{lrrl}\hline
Chart metrics			& Affected	& Percentage	& Base metric	\\ \hline

Charts				& 115		& 100.00	& --		\\
Irregularity: no maintainer	& 95		& 82.61		& Charts	\\
Irregularity: name collision	& 2		& 1.74		& Charts	\\
Irregularity: multiple versions	& 0		& 0.00		& Charts	\\ \hline
\end{tabular}
\end{table}

The Helm \texttt{lint} command, whose task is to catch chart-specific issues, differentiates
between the issue levels \texttt{info}, \texttt{warning} and \texttt{error} with increasing severity.
The command reports zero errors, zero warnings and 20 infos, one per chart, affecting 11.30\% of all charts
on KubeApps Hub.
The infos strictly relate to conventions, pointing out the recommendation to include a graphical icon.
No irregularity is detected.
For comparison, the command reports 37 severe errors, 3 warnings and 109 infos in the GitHub charts,
affecting 89.34\% of the respective charts, confirming the general but not absolute benefit of a quality gate as provided
by KubeApps Hub.

Additional metrics are available from the templates contained within the charts on KubeApps Hub.
55 charts (31.07\%) are variable,
which means that there is at least one variable template value such as an auto-generated password. In total, there
are 428 such variables with non-uniform distribution across all charts.
Furthermore, 865 keys in total appear with duplicated values within the scope of each chart, an average of 4.89 per chart, but
again with non-uniform distribution. The resulting total number of duplicate values is 3784, an average of 4.37 per key and 21.38 per chart.
The global maximum number of duplicate values is 39 whereas the average maximum per chart is 5.95.
Among the most-occurring values are \texttt{httpd-data}, \texttt{TPC}, \texttt{spinnaker-config}, \texttt{pachyderm} and \texttt{httpd-data-nonpersistent}.

\begin{figure}[h]
\centering
\caption{Number of charts per number of variable and duplicate values on KubeApps Hub}
\label{fig:vardupes}
\includegraphics[width=0.8\columnwidth]{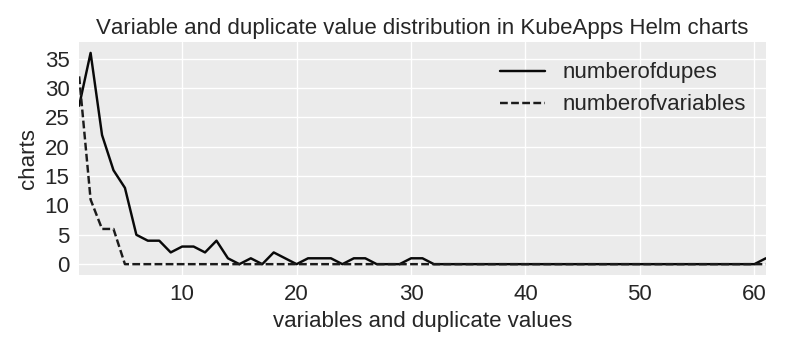}
\end{figure}

Fig \ref{fig:vardupes} overlays the long-tail distributions of variable and duplicate values.
For comparison, the GitHub sample contains 4 variable charts (3.48\%) with 46 variables,
91 keys (average 0.79) with 419 total duplicates (4.60 per key, 3.64 per chart), which is again not consistently
worse than KubeApps Hub.

\subsection{Implications}

Apparently, the irregularities in Helm charts do not appear to impede the popularity of this technology,
and one may question the need to perform more thorough quality assurance. Yet according to the empirical research literature,
software artefact quality matters and if explicitly asked, developers agree or even strongly agree that quality assurance tools
are useful \cite{DBLP:journals/iee/LuciaFST11}.
The value of stricter quality checks will increase in compositions due to the prevention of excessive error propagation.

The absence of maintainer information is particularly severe for our study considering that we aim to
contact maintainers with suggestions for improvements, but may also lead to long-term technical debt
in practice in case bugs are found in dependency charts without the possibility to request bugfixing from
a responsible person.

\section{Methodology}

In order to directly improve the quality of Helm charts published at KubeApps and indirectly those published at other locations,
a methodic approach assisted by the HelmQA tool is chosen. The methodology includes regularly monitoring
the change rates in the repository, automatically generating change suggestions, e-mailing suggested changes to maintainers,
and continuous monitoring of the quality levels to assess the acceptance of the change requests.

\subsection{Hypotheses}

Our software tool-supported empirical research is led by two research questions and two associated hypotheses which are to be confirmed or rejected.
The questions and hypotheses have been pre-registered with the Open Science Framework to ensure the exclusion of bias
by the collected data\footnote{OSF registration site: \url{https://osf.io/u2wn8/register/565fb3678c5e4a66b5582f67}}.
For reasons of practicality, minor deviations from the preregistration are as follows: The data sampling is changed to
once per 24 hours instead of once per 12 hours, and the number of Helm charts is increased from 80 to 177 by the beginning
of the study period in line with the preregistered estimation in the sample size rationale.

The research questions are formulated according to the following excerpt from the preregistration:
\textit{Helm charts are a recent technique to package and configure complex hosted software
applications. Due to manual authoring, the template mechanism to deduplicate congiguration values is often not properly used.
This study investigates two research questions: Can the deduplication be automated with an acceptable amount of false positives?
Will such automation be accepted by software developers and lead to a rapid reduction in duplicate values over months?}

The hypotheses are thus:

\begin{enumerate}
\item If an automation is possible, the ratio of false positives (configuration values marked as duplicated
while they are semantically distinct) will be less than 5\%.
\item If automation is accepted by developers by way of e-mailing them the suggested changes, then at least 50\% of all Helm charts will be
updated within a single month, and the average number of duplicates per chart within this will be reduced by at least 50\%.
\end{enumerate}

\subsection{Algorithms}

Several custom algorithms had to be designed and implemented in HelmQA to measure charts and track changes over time.
They are briefly explained here specifically to rigorously define how the
chart quality, consistency and development metrics are determined.

\paragraph{Version updates.}

Charts are considered related, and different only in version numbers, if the mangled chart name stem is equal. The stem
encompasses all dash-separated name components not starting with a digit and not following one starting with a digit,
and the mangling removes all intermediate dashes
for increased compatibility with graph representations.
For example, \texttt{magic-namespace-0.1.0.tgz} and \texttt{magic-namespace-0.1.1-2.tgz} both result in the common
stem \texttt{magicnamespace}.

A version update (\textit{vupdate}) is recognised if a chart is removed while within the same observation period a chart with
the same stem but different version number is added. This action is distinct from just additions and removals, while
also being different from just updates to the same chart. Version number semantics are not considered as no semantics
are defined for Helm charts.

\paragraph{Variable values.}

While Helm charts contain YAML templates, the analysis needs to take the resulting rendering into YAML files into
account. The rendering is not guaranteed to be stable due to the ability to generate values randomly. This mechanism
is typically used for auto-generated passwords but also directory names.
These variable values represent a challenge when tracking the evolution of the charts due to the resulting diff noise.
Hence, a light-weight feedback-driven machine learning approach is employed as follows: During each invocation of the rendering,
two different outputs are produced and compared. All values which differ are appended to a knowledge base. For subsequent
renderings, these override values are injected into the renderings in a post-processing step.

\paragraph{Duplicate values.}

Similar to variable values, duplicate values may occur in Helm templates whereas they should ideally only occur in the
rendered output through deduplicated variable substitution.
The duplicate check determines the canonical values for all keys in the template and counts the occurrences while skipping
over a blacklisted set of commonly used values including the mandatory markers \texttt{v1} and \texttt{extensions/v1beta1} for Kubernetes objects.
Moreover, occurrences below a threshold, commonly three, are eventually ignored, and the remainder is determined as significant
list of duplicates.

\subsection{Experiment setup}

We re-use the HelmQA scripts developed for the preliminary analysis and run them in an automated way on a nightly basis over approximately four months.
The first month serves as testing and calibration period in which the wellfunctioning of the scripts as well as the baseline indicators
for regular changes to chart files are established. After the first month, another script is run which produces and sends e-mails to
all listed maintainers to inform them about suggested changes. The second month then serves as observation period for quality improvements
likely triggered by the employment of HelmQA. It is expected that the rate of
improvements be significantly above the baseline rate. The third and fourth month serve as long-term continuation and countercheck if the improvement
rate converges back to the baseline rate.
The timeframe for the experiment has been set to mid-May to mid-September 2018.

In order to facilitate the long-term automation of the quality framework and to attract chart maintainers during the observation period, a HelmQA web frontend has been created which dynamically generates
web pages from JSON-structured reports\footnote{HelmQA reports web frontend: \url{http://helmqa-zhaw-hendu.appuioapp.ch/}}. The web frontend implementation along with the core package of HelmQA takes around 1200 significant code lines in Python. Furthermore, a developer-focused website is available which introduces the tool with documentation and
explanation videos\footnote{HelmQA software website: \url{http://serviceprototypinglab.github.io/helmqa/}}.

\section{Results}

\subsection{Experiment execution}

\paragraph{Calibration period (May to June 2018).}

The regular activity encompasses a mean average of 2.73 daily changes to charts, led by new versions of existing
charts (6.33) and updates to existing versions (3.27).
Mentionable is the difference between weekdays with an average of 3.48 daily changes versus weekends with 0.00
updates. The weekdays further include selected holidays including Pentecost and Memorial Day whose reach correlates strongly
with the assumed workplaces of the maintainers under observation.

\begin{figure}[h]
\centering
\caption{Baseline change rates of Helm charts at KubeApps Hub}
\label{fig:baseline}
\includegraphics[width=0.49\columnwidth]{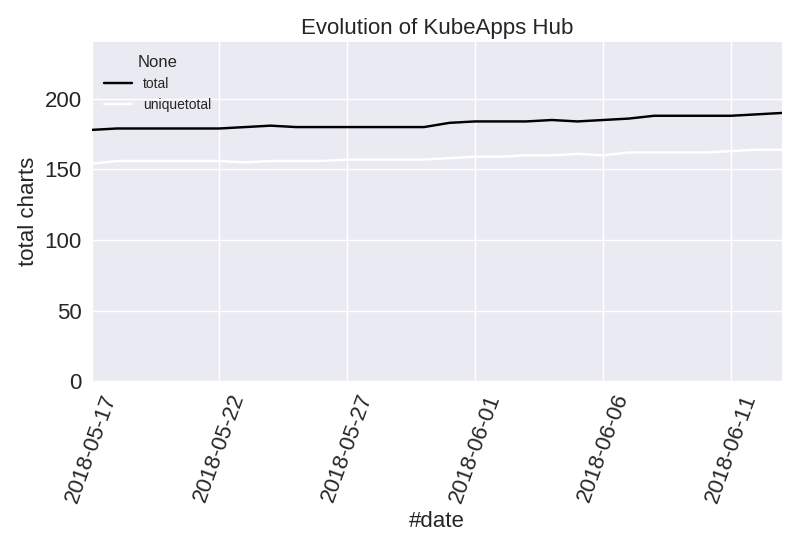}
\includegraphics[width=0.49\columnwidth]{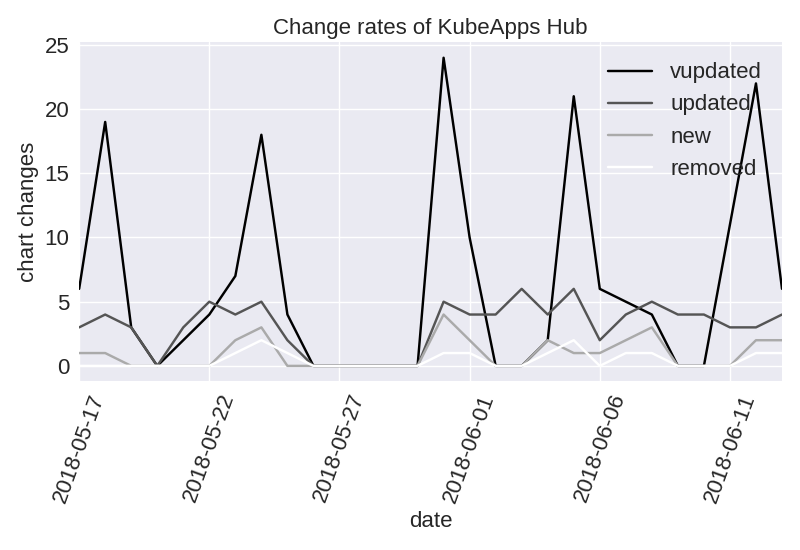}
\end{figure}

Fig. \ref{fig:baseline} shows the change rates for the first-month calibration period. The number of charts increased by 7.3\% to 190
while the versioning overhead only saw a slight increase of 0.2\% to 15.9\%.

According to the gathered
metrics, charts can be clearly clustered into three activity levels: regular changes, infrequent changes and unchanged charts.
Table \ref{tab:baseline} summarises the corresponding metrics for the set of charts present at the beginning of the calibration period.
The causes for the different activities are not further analysed
here. A probable explanation for the most frequent level would be integration with automated continuous build systems.
The two charts in this level are \texttt{anchorage-engine} and \texttt{janusgraph}, both with different sets of maintainers,
followed by the infrequently changed \texttt{spinnaker}, \texttt{redis}, \texttt{redmine} and \texttt{prometheus},
each of which is still changed on at least every third day on average.

\begin{table}[h]
\centering
\caption{Clustered change rates across all Helm charts at KubeApps Hub; DCR = percentage of days with changes}
\label{tab:baseline}
\begin{tabular}{llrrr}\hline
Activity level		& Condition DCR		& Actual DCR	& Charts	& Percentage	\\ \hline

Reglarly changed	& $>50$			& 82		& 2		& 1.13		\\
Infrequently changed	& $>0$ and $<=50$	& 9		& 108		& 61.02		\\
Unchanged		& $0$			& 0		& 67		& 37.85		\\ \hline
\end{tabular}
\end{table}

\begin{figure}[h]
\centering
\caption{Number of e-mail addresses per number of issues}
\label{fig:issues}
\includegraphics[width=0.8\columnwidth]{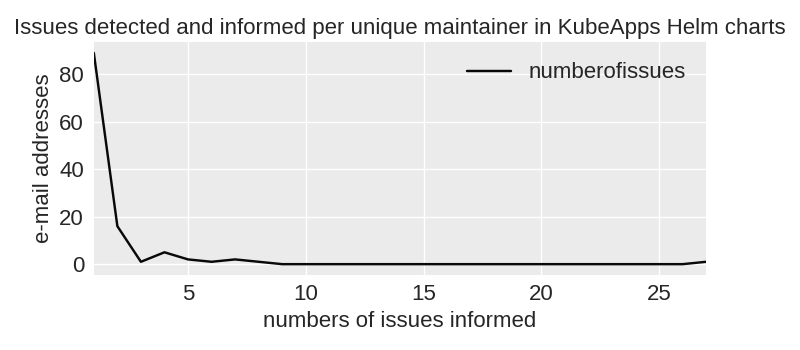}
\end{figure}

At the end of the calibration period, notifications were sent in batches to 118 distinct e-mail
addresses with a total number of 209 unique chart and maintainer issues, on average 1.77 per recipient.
Fig. \ref{fig:issues} shows the long-tail distribution of e-mail addresses links to the number of issues.
Notable is one maintainer with 27 issues, \texttt{Bitnami}/\texttt{bitnamibot}, which coincides with the
highest number of maintained package per maintainer set.

\paragraph{Observation period (June to July 2018).}

The number of recipients reading the notification and actually acting on it is not known.
However, two maintainers reacted on the microblog announcement of HelmQA and three more maintainers by e-mail pointing out
their interest for the idea as well as one technical flaw with the HelmQA web frontend hosting which was quickly
corrected. To gain more insight, the web server logs were analysed. Due to occasional rotation, not all logs
were saved, but at least 25 auto-generated diffs were downloaded by maintainers, including six for multiple
times, suggesting that maintainers checked again after applying changes. While web crawlers and other bots
accessing the diffs cannot be ruled out, this factor can be considered unlikely due to the timing and the
communication of the website links only in private e-mails.

Fig. \ref{fig:observationchanges} shows the resulting change rates for the second-month observation period.
The number of charts increased by another 9.5\% to 208 while the versioning overhead decreased by 1.6\% to 14.3\%.
Whether the removal of duplicate versions were triggered by HelmQA reports is not clear but given the previous
increase a quality-improving influence is likely.

\begin{figure}[h]
\centering
\caption{Effected change rates of Helm charts at KubeApps Hub}
\label{fig:observationchanges}
\includegraphics[width=0.49\columnwidth]{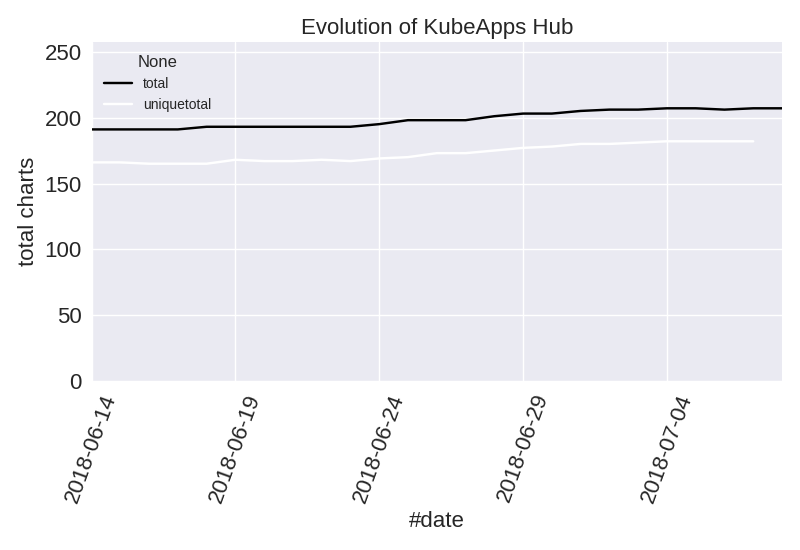}
\includegraphics[width=0.49\columnwidth]{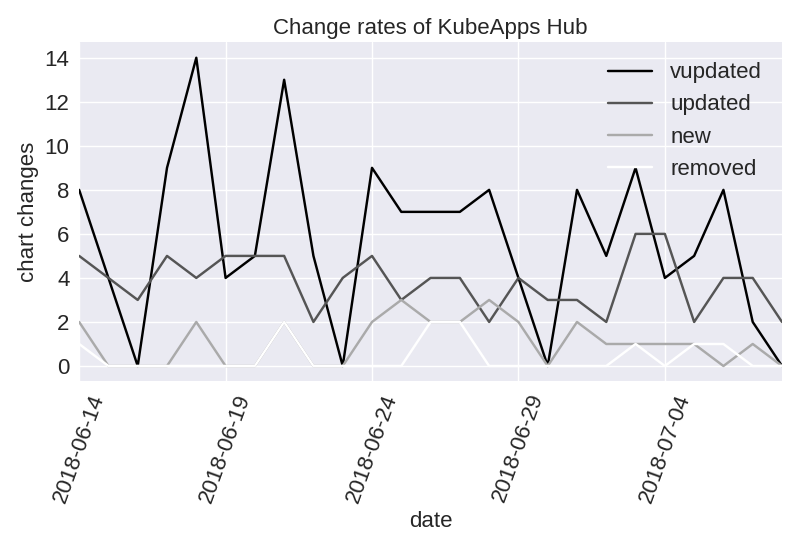}
\end{figure}

\paragraph{Continuation period (July to September 2018).}

The last period was characterised by a refactoring of the HelmQA software so that, while the key functionality would remain,
it would become more attractive to developers compared to only for research tasks. Among the major changes are a modularisation,
proper packaging as Python library, livecheck functionality which can be integrated into continuous development and integration
pipelines (e.g. using Travis or Jenkins) to assess the quality of Helm charts stored in arbitrary repositories,
and a developer-oriented website. In parallel, monitoring of KubeApps Hub continued.

Fig. \ref{fig:continuedchanges} shows the continued change rates for the third- and fourth-month observation period.
In the third month, there was again a growth in the number of charts, albeit slightly reduced, of 7.2\% to 222 in total whereas the versioning
overhead remained almost stable at 14.4\%. In the fourth month, growth slowed even more to 0.9\%, leading to 224 charts with however
an increased versioning overhead of 17.3\%.

\begin{figure}[h]
\centering
\caption{Continued change rates of Helm charts at KubeApps Hub}
\label{fig:continuedchanges}
\includegraphics[width=0.49\columnwidth]{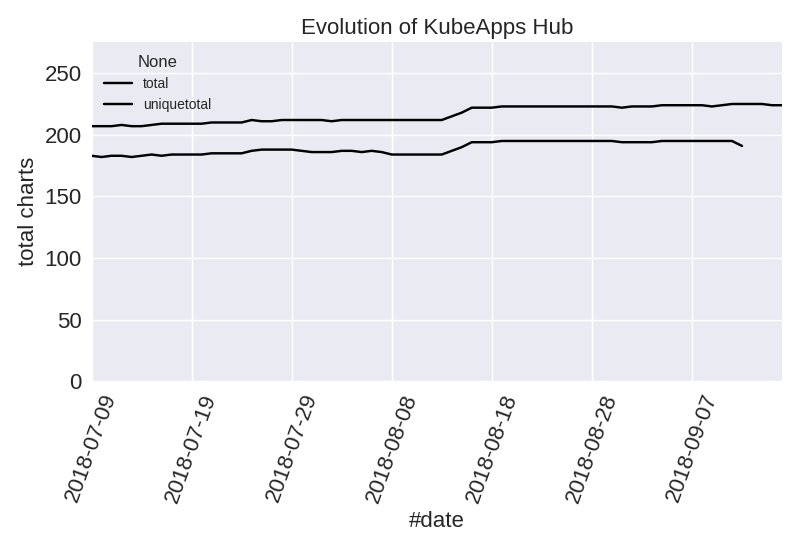}
\includegraphics[width=0.49\columnwidth]{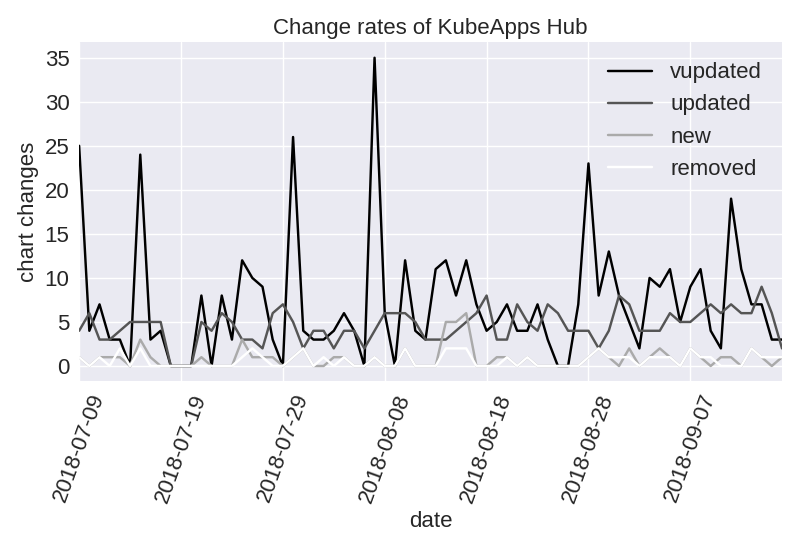}
\end{figure}

\subsection{Improvements and Trends}

Table \ref{tab:periods} summarises the key metrics over the three experiment periods calibration (one month), observation (one month)
and continuation (two months), normalised to per-month numbers. The changed charts encompass all forms of regular and infrequent
updates including \textit{vupdates} compared to the overall number of non-unique charts. Interestingly, in the decisive observation
period, the smallest percentage of charts received updates.

\begin{table}[h]
\centering
\caption{Metric evolution over all experiment periods, monthly average}
\label{tab:periods}
\begin{tabular}{lrrrr}\hline
Metric			& Calibration	& Observation	& Continuation	& Average	\\ \hline

Charts			& +7.3\%	& +9.5\%	& +4.1\%	& +6.8\%	\\
Unique charts		& +0.2\%	& -1.6\%	& +1.5\%	& +0.4\%	\\
Changed charts		& 53.4\%	& 50.5\%	& 50.7\%	& 51.3\%	\\ \hline
\end{tabular}
\end{table}

Going into greater level of detail, Tables \ref{tab:maintainernew} and \ref{tab:chartnew} continue the quality analysis
of Helm charts at KubeApps Hub initially shown in Tables \ref{tab:maintainer} and \ref{tab:chart}. All metrics are calculated
by the 'authorsets' functionality of HelmQA and are absolute irrespective of being proportional to the number of
maintainers or charts. Omitted values signal that metrics remained unchanged.

\begin{table}[h]
\centering
\caption{Maintainer-related KubeApps Hub metrics evolution, monthly average}
\label{tab:maintainernew}
\begin{tabular}{lrrrr}\hline
Maintainer metrics		& Calibr.	& Observ. 	& Continuation	& Average	\\ \hline

Maintainers			& +4.2\%	& +8.8\%	& +4.0\%	& +5.6\%	\\
Sets of maintainers		& +6.2\%	& +10.8\%	& +3.8\%	& +6.6\%	\\
Avg charts per maintainer	& +2.4\%	& +0.8\%	& --		& +0.8\%	\\
Avg charts per maintainer set	& +0.6\%	& -1.3\%	& +0.3\%	& --		\\
Max charts per maintainer set	& --		& --		& --		& --		\\
Avg maintainers per set		& -2.4\%	& -1.6\%	& +0.4\%	& -0.8\%	\\
Max maintainers per set		& --		& --		& --		& --		\\
Irregularity: alias names	& --		& --		& +6.3\%	& +3.1\%	\\ \hline
\end{tabular}
\end{table}

The decisive observation period indicates that presumably due to reports by HelmQA, the usual growth in alias names and in
missing maintainer metadata was suppressed, whereas in contrast, the irregularity of multiple versions of a chart
being present even aggravated.

\begin{table}[h]
\centering
\caption{Chart-related KubeApps Hub metrics evolution, monthly average}
\label{tab:chartnew}
\begin{tabular}{lrrrr}\hline
Chart metrics			& Calibr.		& Observ.	& Continuation	& Average	\\ \hline

Charts				& +7.3\%		& +9.5\%	& +4.1\%	& +6.8\%	\\
Irregularity: no maintainer	& +20.0\%		& --		& --		& +5.0\%	\\
Irregularity: name collision	& --			& --		& --		& --		\\
Irregularity: multiple versions	& $\downarrow$-20.0\%	& +20.0\%	& --		& --		\\ \hline
\end{tabular}
\end{table}

Complementary to the maintainer and chart analysis, Table \ref{tab:templates} informs about the evolution
of template metrics in proportion to the overall number of charts.
These include the ratio of charts with at least one variable or duplicate value in a template to all charts and
the intensity in terms of variables or duplicates per variable or duplicate chart.
One can observe that in the decisive observation period, the irregularity reduction and thus the
quality gain related to duplicates was strongest among all periods.
In some cases, severe template rendering failures mostly due to missing password entries occur;
at the end of the continuation period, this affected templates in 3 charts or 1.3\% of all charts.

\begin{table}[h]
\centering
\caption{Template-related KubeApps Hub metrics evolution, monthly average}
\label{tab:templates}
\begin{tabular}{lrrrr}\hline
Template metrics		& Calibr.	& Observ.		& Continuation	& Average	\\ \hline

Variable charts ratio		& +2.6\%	& -2.2\%		& -0.2\%	& --		\\
Variable charts intensity	& +3.8\%	& -1.2\%		& --		& +0.6\%	\\
Duplicate charts ratio		& -0.7\%	& $\downarrow$-2.1\%	& +0.7\%	& -0.4\%	\\
Duplicate charts intensity	& +1.9\%	& $\downarrow$-3.2\%	& +3.3\%	& +1.3\%	\\
Unrenderable template		& --		& +1.0\%		& +0.3\%	& +0.3\%	\\ \hline
\end{tabular}
\end{table}

\subsection{Findings}

The study results lead to answers to the research questions which are valuable in practice.
Yes, the deduplication can be automated with no evident false positives, but no, the automatically rewritten templates are not directly
accepted by chart maintainers or other developers. More concretely, in the most recent set of KubeApps Hub stable charts,
no occurrence of HelmQA-produced variables of the form \texttt{.suggestions.varN} could be found in any template file. In cases when maintainers
were inspired by HelmQA's generated diffs, they chose a syntactically differently formulated deduplication. However,
while the total ratio of updated (including version-updated) Helm charts within the one-month observation period was slightly above the
anticipated half with 50.5\%, this change activity was rather low and the average number of duplicates was only reduced by 3.2\%
which is far from the anticipated half.
With these results, hypothesis 1 is accepted but hypothesis 2 is rejected.

In statistical terms, according to the preregistered interference criteria, the Wilcoxon signed-rank test over the chosen study periods
is expected to give a p-value of 10\% to determine whether the suggestions to developers are significant for quality improvements.
To determine the value, the entire set of unversioned charts is divided into two groups: the ones which entered the observation period with maintainer notifications ($n_1$)
and the ones which have been added afterwards at the end of the observation period ($n_2$). Then, both sets are compared after the
continuation period, excluding charts which are removed in either set in between.
The resulting set sizes are $|n_1| = 160$ (171 versioned chart files, 149 (87\%) with duplicates) and $|n_2| = 16$ (without additional versioned chart files, 11 (69\%) with duplicates),
with $n_2$ containing more recent charts such as \texttt{tomcat} and \texttt{heartbeat}.
The median number of duplicate keys in $n_1$ considering all versions at the end of the continuation period is 5.15, and in $n_2$ it is 2.75.
The smallest p-value over equal-sized samples of size $n = n_2$ after 1'000'000 independently sampled tests is 0.04\%.
The influence of the chosen suggestion and notification procedure on the quality is therefore statistically insignificant although a larger sample size of $n_2$
would be required to ensure the validity of the test.

\section{Discussion}

Wider technical and social implications are revealed by our study, both chronologically for future Helm charts and spatially for other ecosystems. Quality checks need to become stricter in tools and marketplaces, but there will always be external tools to provide additional checks. Hence, digital artefact repositories need an extensible tagging and filtering mechanism so that their contents can be tagged after successful external quality assessment and developers can filter out all artefacts without the required tags. For Helm charts in particular, regular reminders to maintainers and a prior assurance that the reminders can be properly addressed are expected to lead to long-term quality improvements. Interpolating from the current growth trend, we can assume that by the end of 2019, there will be around 400-450 charts in KubeApps Hub which, due to the network effect, will imply an at least three-fold popularity among composite cloud application developers compared to the begin of our study, fueling further charts who build upon the additionally available dependencies. With the increased use of Helm charts for business-critical deployments, the need for quality assessment and assurance will increase.

Although auto-generated code is common in software development, auto-tuned or modified code appears to be accepted only at runtime, not at implementation time with the exception of syntax formatters (pretty printers). Yet deduplication of values is important at the implementation stage precisely because quality errors may result from missing selected updates. Our suggestion is that integrated development environments convey quality assessment of cloud application artefacts to remove the barrier between suggestions and actual fixes, following recent studies which indicate that the context and priorisation of such information is important \cite{DBLP:conf/wcre/VassalloPPPZG18}.

\section{Related Work}

Researchers have contributed various statistics about digital artefact ecosystems although in some cases no software tools are provided to re-run the experiments. In most cases, the available tools only address researchers and are not targeting developers. Among the cloud artefacts, most studies focus on Docker Hub and its vast repository of Docker container images.
Ayed et al. provide a SPARQL query interface called DockerWebStats based on Docker2RDF to enable research on container image metadata \cite{DBLP:conf/services/AyedSLCLK17}. As early research, it was produced when only 120 official images were available on Docker Hub in December 2016, and as of September 2018 the web application is no longer available on the specified IP address.
Shu et al. have performed a security vulnerability analysis of more than 350'000 community images and, indicating a rapid growth, more than 3'800 official images \cite{DBLP:conf/codaspy/ShuGE17}. Shu used the externally developed Clair tool which in 2018 is still actively maintained by CoreOS, now Red Hat, and only focuses on security.
Cito et al. studied the quality level of more than 70'000 Dockerfiles on GitHub \cite{DBLP:conf/msr/CitoSWLZG17}. In contrast to Helm charts which statically include dependencies in the form of other (small) charts, Docker images are large and use a referencing mechanism. Hence, Cito found that among the most frequent quality issues were missing version pinning, affecting more than 28\% of all files, along with frequent build failures estimated at more than 30\%. This corresponds to Helm charts whose templates do not render which according to our study affects 1.3\%.

\section{Conclusion}

\subsection{Summary}

We have contributed the HelmQA software tool which enables cloud application developers, especially maintainers of Helm charts for the
Kubernetes application ecosystem, to assess the quality of declarative chart artefacts. The tool is typically integrated into CI/CD
pipelines through its livecheck functionality or occasionally run to get information about a single chart of set of charts.
In our study, we have continuously applied the tool over the growing set of charts maintained at KubeApps Hub and actively
informed maintainers about quality deficiencies. Surprisingly, while a few metrics indicate that the automatically generated
improvement suggestions have been picked up by the maintainers, the overall level of improvement is small with 3.2\% less duplicate
values in templates as well as 0\% less irregularities concerning lack of maintainer information, name collision and alias names.
We foresee the enforced use of the tool or equivalent functionality as part of the Helm linting process and the KubeApps Hub submission
process as only way to ensure consistently higher quality compared to the current state and to the decentrally hosted charts.

\subsection{Reproducibility}

The observed evolution of Helm charts at KubeApps Hub is a unique historic dataset which we provide for reference at the
associated Open Science Framework website\footnote{HelmQA datasets: \url{https://osf.io/5gkxq/}}.
However, similar observations can be expected by reproducing our experiments, including a regular retrieval of KubeApps Hub
charts, automated invocation of HelmQA checks, and interaction with chart maintainers.

\bibliographystyle{spbasic}
\bibliography{draft}

\end{document}